# Graph Computing based Distributed State Estimation with PMUs

Yi Lu[a], Chen Yuan[b], Xiang Zhang[b], Hua Huang[c], Guangyi Liu[b], Renchang Dai[b], Zhiwei Wang[b]

[a] State Grid Sichuan Electric Power Company, Chengdu, Sichuan, China
[b] GEIRI North America, San Jose, CA, USA
[c] NARI Technology Co., Ltd
luyi_1230@126.com

*Abstract*—**Power system state estimation plays a fundamental and critical role in the energy management system (EMS). To achieve a high performance and accurate system states estimation, a graph computing based distributed state estimation approach is proposed in this paper. Firstly, a power system network is divided into multiple areas. Reference buses are selected with PMUs being installed at these buses for each area. Then, the system network is converted into multiple independent areas. In this way, the power system state estimation could be conducted in parallel for each area and the estimated system states are obtained without compromise of accuracy. IEEE 118-bus system and MP 10790-bus system are employed to verify the results accuracy and present the promising computation performance.**

## I. Introductoin

System state estimation (SE) is a fundamental application in the energy management system (EMS) to estimate power system states via measurements [1]. In current practice, SE runs every 1–5 minutes for large power systems, which leads to a 1–5 minutes delay between the current power system and the estimated one. If a severe event happens, there may be a large difference between the estimated system states and the true real-time system states. It is very hard for system operators to identify the problem and secure the system in a timely manner. Moreover, with the increased complexity of power systems due to the penetration of distributed/renewable energy resources [2]–[4], electric vehicles [5], responsive loads [6], [7], and even potential cyber-attacks [8], more frequent—and even rapid—changes in system states have been introduced. The U.S. Department of Energy (DOE) has presented the computational needs for next-generation electrical grids in [9]. It states that the future direction for SE is to reduce the solution time from minutes to seconds—or even milliseconds. This would allow operators to monitor the system states in real-time and to secure the system's operation in a timely manner. Furthermore, a fast SE could also help expedite other network analysis applications and speed up the EMS.

The idea of multiprocessor SE was proposed in [10], [11]. In [10], distributed processors were employed for WLS SE, where local results were transmitted to a central processor once they were received by a substation processor. [11] reordered the system nodes for parallel computing before forward/backward substitution (FBS). In addition, previous work in [12], [13] discussed distributed SE. In both, the system network was partitioned into several areas, and in each area, there was a local control center. Then the local SE results were obtained before they were coordinated to get the results. The measurements on the boundary were ignored due to the structure of the multi-area system. The results' accuracy was sacrificed.

In this paper, a graph computing based distributed state estimation approach is proposed to speed up the computation performance without the compromise of results accuracy. At first, a power system network is divided into multiple areas. Boundary buses on the terminals of inter-area lines are selected as reference buses with PMUs being equipped at these buses. At each sampling period, the voltage magnitude and phase angle of each reference bus are recorded from PMUs. Then, the system is transformed into multiple independent areas. In this way, the state estimation could be conducted in parallel and independently for each area and the system states could be guaranteed without the compromise of accuracy. In addition, graph computing based WLS fast decoupled state estimation is employed to quickly calculate system states for each area.

This paper is organized as follows. In Section II, the graph computing and system partitioning are briefly introduced. Then, the graph computing based distributed FDPF method is elaborated in Section III. Case study is conducted in Section IV to verify the results accuracy of the proposed approach and demonstrate its significant computation efficiency. The conclusion is summarized in Section V.

## II. Graph Computing and System Partitioning with Phasore Measurement Unit

### A. Graph Database and Graph Computing

Graph is a data structure modeling pairwise relations between objects in a network. In mathematics, a graph is represented as $G = (V, E)$, in which $V$ indicates a set of vertices, representing objects, and the set of edges is

This work is supported by the State Grid Corporation technology project 5455HJ180020

represented as $E$, expressing how these objects relate to each other. Each edge is denoted by $e = (i,j)$ in $E$, where $i$ and $j$ in $V$ are referred as head and tail of the edge $e$, respectively [14].

Previous works have explored the usage and feasibility of the graph database to naturally represent power systems and apply graph computing to energy management systems in power grids [15]. In this subsection, two main parts of graph computing, i.e. node-based parallel computing and hierarchical parallel computing, are presented as follows.

*1) Node-based Parallel Computing*: In graph computing, node-based parallel computing represents that the computation at each node is independent and can be conducted in parallel. In Fig. 1, its upper half depicts the mapping relation between graph-based computation and matrix computation. It clearly shows that, in a graph, the counterparts of the connections between nodes are non-zero off-diagonal elements in the coefficient matrix, $A$. Zero off-diagonal elements in the coefficient matrix indicate that no direct connections between the nodes exist in the graph. The bottom half of Fig. 1 demonstrates the node-based parallel computing strategy. Taking the admittance matrix in Fig. 1 as an example, the off-diagonal element is locally calculated based on the impedance attributes of the corresponding edge, and each diagonal element is independently obtained only with the processing of impedance attributes at the corresponding node and its connected edges. Therefore, the whole admittance matrix is developed with one-step graph operation and the value of each element is calculated independently and in parallel. Other examples of node-based parallel computation in power systems are active/reactive power injection calculation, node variables mismatch update, active/reactive power flow calculation, etc.

*2) Hierarchical Parallel Computing:* Hierarchical parallel computing performs computation for nodes at the same level in parallel. The level next to it is performed after. An example of hierarchical parallel computing application is matrix factorization. Cholesky elimination algorithm is employed to conduct matrix factorization. Three steps are involved: 1) determining fill-ins, 2) forming an elimination tree, and 3) partitioning the elimination tree for hierarchical parallel computing.

### B. System Partitioning for Distributed Graph Computing

In the previous subsection, the concept of graph computing is presented. But, as it depicts, it takes the entire system as an entity and conducts the processing together. All the nodes do the local computing first, then communicate and convey information with others, and at last wait for synchronization.

To further speed up the power flow analysis performance, system partitioning, considering line cutting minimization, geolocation, voltage level of transmission line, etc., is used for distributed graph computing. Since the paper mainly focuses on implementing distributed power flow analysis with developed system partitioning [12], [16], the exact partitioning approach is out of scope. After the system partitioning, the overall system is decomposed into multiple non-overlapping areas, so that each area could be processed independently and in parallel using graph computing technique. In other words, each area is processed independently, and, within each area, graph computing is applied to implement parallel processing. So, the burden of local computation, communication, and synchronization for each area is much reduced since the dimension of each area is reduced, as shown in Fig. 1.

## III. GRAPH COMPUTING BASED DISTRIBUTED STATE ESTIMATION WITH LIMITED PMUS

In this section, the approach of distributed graph computing will be applied to power system state estimation and its details are illustrated.

### A. Graph Computing based WLS Fast Decoupled State Estimatoin

For an $n$-bus power system with $m$ measurements, its nonlinear WLS SE problem is formulated as [17]:

$$\min\ J(x) = [z - h(x)]^T R^{-1}[z - h(x)] \quad (1)$$

where $z$ is in the dimension of $m$; $x$ is in the dimension of $2n - 1$; $h$ is the nonlinear function of $x$, which relates the system states to the error-free measurements; and $R^{-1}$ is a diagonal matrix, consisting of the weight, $\frac{1}{\sigma_j^2}$, for each measurement $j$.

To achieve the minimum of $J(x)$, the equation below is derived:

$$g(x) = \frac{\partial J(x)}{\partial x} = -H^T(x)R^{-1}r(x) = 0 \quad (2)$$

where $H(x) = \frac{\partial h(x)}{\partial x}$, and $r(x) = z - h(x)$.

Substituting the first-order Taylor expansion of $g(x)$ in (2), the solution of the objective function can be found by iteratively solving (3).

$$G(x^k)\Delta x^k = H^T(x^k)R^{-1}r(x^k) \quad (3)$$

where $G(x^k) = \frac{\partial g(x^k)}{\partial x} = H^T(x^k)R^{-1}H(x^k)$, $x^{k+1} = x^k + \Delta x^k$, and $x^k$ is a vector of system states at iteration $k$.

Assuming a high X/R ratio in power transmission systems, the fast decoupled method is employed to convert the $H$ matrix and gain matrix in (3) to constant matrices, thus saving time on repetitive matrix formulation and factorization [18]. Besides, in (4) and (5), the numerical values corresponding to the off-diagonal blocks of matrix $H$, and consequently those of gain matrix, are significantly smaller than those of the diagonal blocks, so the non-diagonal blocks are neglected. The subscripts $A$ and $R$ denote the corresponding active and reactive components, respectively.

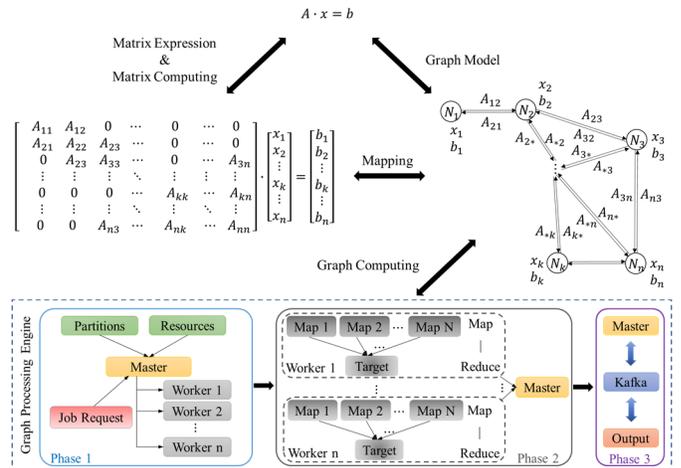

Figure 1. Mapping between graph computation and matrix computation

$$H = \begin{bmatrix} H_{AA} & H_{AR} \\ H_{RA} & H_{RR} \end{bmatrix} = \begin{bmatrix} H_{AA} & 0 \\ 0 & H_{RR} \end{bmatrix} \quad (4)$$

$$G = \begin{bmatrix} G_{AA} & G_{AR} \\ G_{RA} & G_{RR} \end{bmatrix} = \begin{bmatrix} H_{AA}^T R_{AA}^{-1} H_{AA} & 0 \\ 0 & H_{RR}^T R_{RR}^{-1} H_{RR} \end{bmatrix} \quad (5)$$

where, $R^{-1} = diag(R_{AA}^{-1}, R_{RR}^{-1})$. In (5), $G_{AA} \in \mathcal{M}(n-1, n-1)$, while $G_{RR} \in \mathcal{M}(n, n)$. This is because the voltage angle at the swing bus is the reference, so the derivatives of the measurements to the swing bus angle are not included in (4).

Then, equation (3) is rewritten as:

$$\begin{cases} G_{AA} \Delta \theta^k = H_{AA}^T R_{AA}^{-1} r_A(x^k) & (6-a) \\ G_{RR} \Delta |V|^k = H_{RR}^T R_{RR}^{-1} r_R(x^k) & (6-b) \end{cases}$$

Since the approach to solve the (6-a) and (6-b) is the same, the following will use (6-a) as the example to further explain how to implement the WLS fast decoupled state estimation with graph computing technique.

*1) Node-based parallel computing – formulating Jacobian matrix and gain matrix, calculating right-hand-side vector and updating system states:* To realize high-performance power system SE, the node-based graph computing technique is employed to efficiently formulate the problem in the system graph model. Detailed derivations are elaborated as follows.

At first, system measurements are collected and assigned into node-based, *i.e.* bus-based, measurement parts, and measurements in the same part are directly related to the same bus. For an *n*-bus system, the vector of active system measurements, $z_A$, is first reordered by putting each bus's directly related measurements together as a measurement sub-vector and sorting each sub-vector per bus index. Then $z_A$ is partitioned into $n$ parts, and each partition belongs to a distinct bus. The whole process is further presented in (7)-(9). Without loss of generality, it is assumed that the system has $n$ buses. In (7), the system measurement vector, $z_A$, is partitioned into $n$ parts. $z_{A,i}$ indicates the bus *i*'s measurement sub-vector and, as presented in (8), it includes bus *i*'s active power injection measurement, $P_i$, and the vector of active branch power flow measurements at bus $i$, $P_{ij}$. The error-free vector is accordingly divided into n parts in (9).

$$z_A = [z_{A,1}^T \ z_{A,2}^T \ \cdots \ z_{A,i}^T \ \cdots \ z_{A,n}^T]^T \quad (7)$$

$$z_{A,i} = [P_i \ P_{ij}^T]^T, \ i = 1, \cdots, n \quad (8)$$

$$h_A(x) = [h_{A,1}(x)^T \ h_{A,2}(x)^T \ \cdots \ h_{A,i}(x)^T \ \cdots \ h_{A,n}(x)^T]^T \quad (9)$$

Then, the $H$ matrix and gain matrix of bus $i$ are derived as:

$$H_{AA,i} = \frac{\partial h_{A,i}(x)}{\partial \theta}, \ \theta = [\theta_1 \ \theta_2 \ \cdots \ \theta_i \ \cdots \ \theta_{n-1}] \quad (10)$$

$$G_{AA,i} = H_{AA,i}^T R_{AA,i}^{-1} H_{AA,i} \quad (11)$$

where $R_{AA,i}^{-1}$ is the weight matrix for active power measurements directly related to node *i*. $H_{AA,i}$ and $G_{AA,i}$ can be locally calculated at bus *i* with at most one step graph traversal since only information of bus *i*, bus *i*'s 1-step neighbor(s) and the branch(es) between them are stored in $H_{AA,i}$ and $G_{AA,i}$.

The system-level $H$ matrix, $H_{AA}$, is then formulated with the aggregation of all node-based $H$ matrices, $H_{AA,i}$, while the system-level gain matrix, $G_{AA}$, is obtained by summing up the node-based gain matrices, $G_{AA,i}$. The detailed derivations are presented in (12) and (13).

$$H_{AA} = \frac{\partial h_A(x)}{\partial \theta}$$

$$= \frac{\partial [h_{A,1}(x)^T \ h_{A,2}(x)^T \ \cdots \ h_{A,i}(x)^T \ \cdots \ h_{A,n}(x)^T]^T}{\partial \theta}$$

$$= \left[ \left(\frac{\partial h_{A,1}(x)}{\partial \theta}\right)^T \ \left(\frac{\partial h_{A,2}(x)}{\partial \theta}\right)^T \ \cdots \ \left(\frac{\partial h_{A,i}(x)}{\partial \theta}\right)^T \ \cdots \ \left(\frac{\partial h_{A,n}(x)}{\partial \theta}\right)^T \right]^T$$

$$= [H_{AA,1}^T \ H_{AA,2}^T \ \cdots \ H_{AA,i}^T \ \cdots \ H_{AA,n}^T]^T \quad (12)$$

$$G_{AA} = H_{AA}^T R_{AA}^{-1} H_{AA}$$

$$= [H_{AA,1}^T \ H_{AA,2}^T \ \cdots \ H_{AA,i}^T \ \cdots \ H_{AA,n}^T] \cdot R_{AA}^{-1} \cdot$$

$$[H_{AA,1}^T \ H_{AA,2}^T \ \cdots \ H_{AA,i}^T \ \cdots \ H_{AA,n}^T]^T$$

$$= \sum_{i=1}^{n} H_{AA,i}^T \cdot R_{AA,i}^{-1} \cdot H_{AA,i}$$

$$= \sum_{i=1}^{n} G_{AA,i} \quad (13)$$

where, $R_{AA}^{-1} = diag(R_{AA,1}^{-1}, R_{AA,2}^{-1}, \cdots, R_{AA,i}^{-1}, \cdots, R_{AA,n}^{-1})$

The system-level RHS vector update in (6) can also be implemented by summing up the node-based RHS vectors, as shown in (14). The update of each node-based RHS vector is locally computed.

$$RHS = H_{AA}^T R_{AA}^{-1} r_A(x^k)$$

$$= [H_{AA,1}^T \ H_{AA,2}^T \ \cdots \ H_{AA,i}^T \ \cdots \ H_{AA,n}^T] \cdot R_{AA}^{-1}$$

$$\cdot [r_{A,1}(x^k) \ r_{A,2}(x^k) \ \cdots \ r_{A,i}(x^k) \ \cdots \ r_{A,n}(x^k)]$$

$$= \sum_{i=1}^{n} H_{AA,i}^T R_{AA,i}^{-1} r_{A,i}(x^k) \quad (14)$$

*2) Hierarchical parallel computing – power system states estimation:* The gain matrix LU factorization and the SE problem-solving via forward/backward substitution are conducted using a graph computing-based high-performance solver developed in the authors' previous work [19], [20].

To summarize the above-elaborated process, the procedure for the proposed graph computing-based WLS fast decoupled SE is presented below.

---

**Graph Computing-based WLS Fast Decoupled State Estimation Procedure:**

1. Start: set iteration index $k = 0$, and initialize the system state vector $x^0$, including $\theta^0$ and $|V|^0$;

2. Formulate gain matrices, $G_{AA}$ and $G_{RR}$, based on node-based graph computing;

3. LU Factorize $G_{AA}$ and $G_{RR}$;

4. Update RHS vector with node-based graph computing;

5. Solve $\Delta\boldsymbol{\theta}^k$, and update $\boldsymbol{\theta}^{k+1} = \boldsymbol{\theta}^k + \Delta\boldsymbol{\theta}^k$;

6. Check convergence: $\|\Delta\boldsymbol{\theta}^k\|_\infty \leq \epsilon_\theta$ and $\|\Delta|\boldsymbol{V}|^{k-1}\|_\infty \leq \epsilon_V$? If yes, output $\boldsymbol{\theta}^{k+1}$ and $|\boldsymbol{V}|^k$; If no, go to step 7;

7. Update RHS vector with node-based graph computing;

8. Solve $\Delta|\boldsymbol{V}|^k$, and update $|\boldsymbol{V}|^{k+1} = |\boldsymbol{V}|^k + \Delta|\boldsymbol{V}|^k$;

9. Check convergence: $\|\Delta\boldsymbol{\theta}^k\|_\infty \leq \epsilon_\theta$ and $\|\Delta|\boldsymbol{V}|^k\|_\infty \leq \epsilon_V$? If yes, output $\boldsymbol{\theta}^{k+1}$ and $|\boldsymbol{V}|^{k+1}$; If no, $k = k + 1$, go to step 4.

### B. Graph Computing based Distributed State Estimatoin

To further improve the performance of power system state estimation, system partitioning into multiple areas will be described in this subsection, and then followed by the proposed distributed state estimation method with PMUs.

In the conventional distributed approach, system network is first divided into multiple areas based on the geological information and topology structure [12]. Then, each area performs its own state estimation and coordinates with others to find the solution by exchanging information at the border. In this paper, the partitioned areas are considered independent from others. Boundary buses on the terminals of inter-area lines are selected as reference buses with PMUs. At each sampling period, the voltage magnitude and phase angle of each reference bus are recorded from PMUs. Then the state estimation in each area is calculated independently. In other words, no more information exchange is needed at the boundary of different areas during the execution of power system state estimation, since the boundary information has been included in the voltage magnitude and phase angles recorded from PMU.

As shown in Fig. 2, IEEE 14-bus system is divided into four areas [12]. There are 7 inter-area branches. Taking branch 4-5 as an example, it is interconnecting area 1 and area 2, linking bus 4 and bus 5. During the process of system network partitioning, branch 4-5 is removed, and buses 4 and 5 are selected as reference buses and equipped with PMUs to collect the measurements of bus voltage magnitudes and phase angles. Similarly, the rest of boundary buses on the terminals of inter-area branches are chosen as reference buses and equipped with PMUs as well. After the inter-area branches removal and reference buses selection with PMUs, the IEEE 14-bus system is partitioned into four isolated areas, as shown in Fig. 2. The buses with dashed circles were once connected with inter-area branches. They are reference buses and have PMUs installed after system partitioning. Since IEEE 14-bus system is a very small system and its admittance matrix is dense, the ratio of impacted buses after system partitioning is high. Here, the impacted buses indicate the boundary buses. But, for larger systems, like IEEE 118-bus system and MP 10790 system, buses are not tightly connected with others and we will see their ratios of impacted buses are very low in Section IV.

After the system network partitioning, the graph computing based WLS fast decoupled state estimation is applied to each area simultaneously and independently.

## IV. PERFORMANCE EVALUATION AND DISCUSSION

### A. Testing Environment and Testing Cases

To verify the results accuracy and demonstrate the high computation performance of the proposed approach, the testing is conducted in a Linux server with the installation of a graph database platform. The detailed configurations of the testing environment are listed below in Table I.

TABLE I. TESTING ENVIRONMENT

| Hardware Environment | |
|---|---|
| CPU | 2 CPUs × 6 Cores × 2 Threads @ 2.10 GHz |
| Memory | 64 GB |
| **Software Environment** | |
| Operating System | CentOS 6.8 |
| Graph Database | TigerGraph v2.3.3 |

In the following subsections, the IEEE 118-bus system, which is a simple approximation of the American Electric Power system in the United States mid-west area, is used to first verify the accuracy of the proposed approach, and MP 10790-bus system [16], which is a system with four interconnected European systems, is employed to demonstrate the promising computation performance.

### B. Approach Verification

Table II provides the results accuracy, using IEEE 118-bus system. The system partitioning of the IEEE 118-bus system is displayed in Fig. 3 [12]. The ratio of impacted buses by equivalent extra load injections is only 11%, indicating that after system partition only 13 buses are impacted, approximately 11% of the total 118 buses, with the addition of an extra equivalent load.

The convergence thresholds for both the voltage magnitude and the phase angle (in radian) are 1.0E-4. The testing is based on flat-start, and the results, including the number of iterations and estimation accuracy, are displayed in Table II, showing good performance in convergence and accuracy. The true values of the system states used in mean squared error calculation are obtained from data file of the IEEE 118-bus

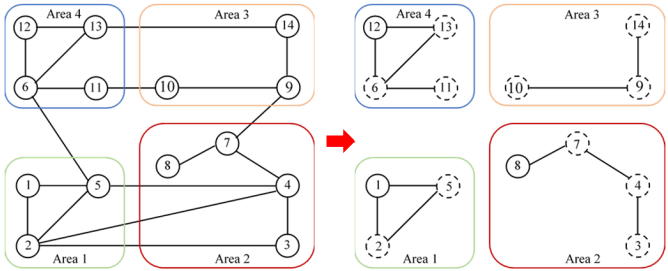

Figure 2. IEEE 14-bus system partitioning

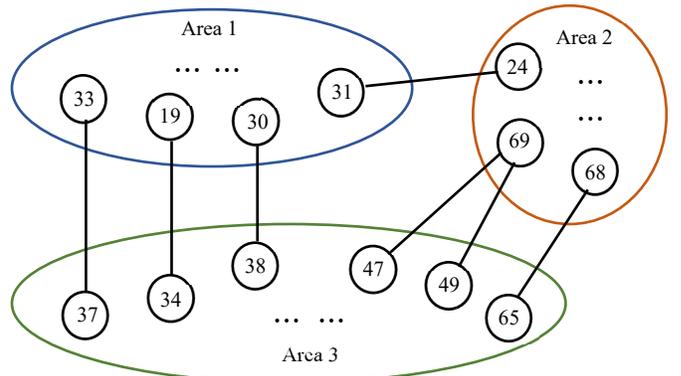

Figure 3. IEEE 118-bus system partitioning

system, which are calculated based on the conventional state estimation method.

Table II. ACCURACY VERIFICATION USING IEEE 118-BUS SYSTEM

| Method | Number of Iterations | Mean Squared Error | |
|---|---|---|---|
| | | Phase Angle (degree) | Voltage Magnitude (per unit) |
| Proposed Method | 5 | 3.75E-9 | 6.78E-14 |

## C. High Computation Performance

After accuracy verification, the testing to demonstrate the proposed method's promising computation efficiency with multiple threads is presented in this subsection. MP 10790-bus system is used to test the parallel computing performance of the proposed method. As displayed in Fig. 4, the system is partitioned into 4 areas [16]. The ratio of impacted buses with equivalent extra load injections is less than 0.06% since only 6 buses are selected as reference buses after system partitioning.

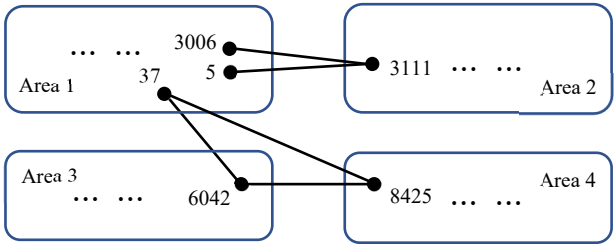

Figure 4. MP 10790-bus system partitioning

The convergence thresholds for both the voltage magnitude and the phase angle (in radian) are 1.0E-4. The testing is based on flat-start, and the testing results, including the number of iterations and computation time, are presented in Table III. Not only the parallelism testing of the proposed method is conducted, but also the computation performance of the graph computing based WLS fast decoupled state estimation method with no partition [21], [22] is presented as a comparison. From Table III, it can be concluded that (a) with the increase of running threads, computation performance is greatly improved, and (b) with the help of system partitioning, the performance of the proposed method is further improved. Regarding the conclusion (a), it demonstrates the promising computation capability of graph computing based approaches. In the conclusion (b), it emphasized the improved efficiency with smaller matrix operation. Besides, the simultaneous power flow analysis for multiple areas largely reduced the time cost.

## V. CONCLUSION

In this paper, a graph computing based distributed state estimation was presented to further speed up the computation performance without compromising results accuracy. Through graph partition, reference buses are selected with PMUs. Then the graph computing based state estimation is applied to each area independently. IEEE 118-bus system and MP 10790-bus system are employed to verify the results accuracy and display the significant computation efficiency of the proposed method.

TABLE III. PARALLEL COMPUTATION PERFORMANCE TESTING WITH MP 10790-BUS SYSTEM

| Method | Number of Iterations | Computation Time under Different Number of Running Threads (ms) | | | |
|---|---|---|---|---|---|
| | | 1 | 2 | 4 | 8 |
| Graph Computing based SE | 7 | 2820.33 | 1544.04 | 978.72 | 641.72 |
| Proposed Method | Area 1: 7 Area 2: 7 Area 3: 5 Area 4: 5 | 2255.43 | 1265.28 | 708.37 | 481.44 |